\documentclass[a4paper,11pt]{article}
\usepackage{jheppub20}

\pdfoutput=1

\usepackage{mathtools}
\usepackage{slashed}
\usepackage{xcolor}
\usepackage{multirow}

\usepackage[normalem]{ulem}
\usepackage{xcolor}

\usepackage{feynmp-auto}

\usepackage{accents}

\usepackage{braket}

\allowdisplaybreaks

\usepackage{siunitx} 
\usepackage{xparse}
\NewDocumentCommand{\fr}{>{\SplitArgument{1}{,}}m}{\efrac#1}
\NewDocumentCommand{\efrac}{mm}{\ensuremath{\frac{#1}{#2}}}

\usepackage{bbold}

\def\hat{\widehat}
\def\tilde{\widetilde}

\title{Quark-sector Lorentz violation in $Z$-boson production}
\author{Enrico Lunghi,$^{a,b}$}
\author{Nathan Sherrill,$^{a,b}$}
\author{Adam Szczepaniak,$^{a,b,c}$}
\author{Alexandre Vieira$^{d}$}
\affiliation{
$^a$Physics Department, Indiana University, Bloomington, IN 47405, USA \\
$^b$Center for Exploration of Energy and Matter, Indiana University, Bloomington, IN 47403, USA \\
$^c$Theory Center, Thomas Jefferson National Accelerator Facility, Newport News, VA 23606, USA \\
$^d$Universidade Federal do Tri\^angulo Mineiro, Campus Iturama, 
38280-000, Iturama MG, Brazil 
}
\emailAdd{elunghi@indiana.edu} 
\emailAdd{nlsherri@iu.edu}
\emailAdd{aszczepa@indiana.edu}
\emailAdd{alexandre.vieira@uftm.edu}

\abstract{
Quark-sector Lorentz violation is studied in the context of Drell-Yan dilepton production including effects from $Z$-boson exchange. We show the chiral nature of the weak interactions enables parity-violating and spin-dependent effects to be studied using unpolarized initial states. Constraints are placed on dimensionless and CPT-even coefficients for Lorentz violation for the first two generations of quarks using measurements from the Large Hadron Collider.}

\begin{document}
\newcommand\redsout{\bgroup\markoverwith{\textcolor{red}{\rule[0.5ex]{2pt}{0.4pt}}}\ULon}
\maketitle

\section{Introduction}
\label{sec:intro}

Collisions of high-energy hadrons continue to serve as laboratories for precision measurements of known physics and searches for new physics beyond the Standard Model (SM) \cite{Proceedings:2019mbn, Brooijmans:2020yij, Proceedings:2020aql}. A subset of these processes, where the production and decay of intermediate bosons occurs, or Drell-Yan (DY) processes \cite{Drell:1970wh}, have been studied for over half a century and have proved to be indispensable in validating the modern theory of strong interactions, quantum chromodynamics (QCD). In particular, the study of DY processes has been essential for understanding the structure of hadrons and establishing features of hadronic collisions; the discovery of the SM electroweak gauge bosons and heavy quarks; and searches for new physics at the Large Hadron Collider (LHC) \cite{Peng:2016ebs}.

An interesting potential signal for new physics that has gained considerable attention in recent years 
is the breakdown of CPT and Lorentz invariance \cite{Proceedings:2020pzb}.
These principles have been tested with exceptional precision in numerous experiments
involving SM particles and gravitational fields 
\cite{Kostelecky:2008ts}. The model-independent framework for 
quantifying tiny deviations from exact CPT and Lorentz invariance, potentially emerging 
from new short-distance physics, is based on effective field theory 
and is known as the Standard-Model Extension (SME) \cite{Colladay:1996iz,Colladay:1998fq,Kostelecky:2003fs,Kostelecky:2020hbb}.
The Lagrange density for the SME incorporates all CPT- and Lorentz-violating terms
for gravitational fields coupled to the SM.
Each term is a coordinate-independent contraction
of a coefficient for Lorentz violation with a product of field operators that transform covariantly under observer Lorentz transformations \cite{Colladay:1996iz}.  
Operators in the SME are classified according to their mass dimension $d$. The finite set of gauge-invariant and renormalizable operators with $d = 3, 4$ in Minkowski spacetime is known as the minimal SME. The infinite set of operators with $d\geq 5$ is referred to as the nonminimal SME. 

Experiments involving hadrons provide opportunities to constrain the comparatively unexamined 
CPT- and Lorentz-violating effects on mesons, baryons, and underlying quark and gluon (parton) degrees of freedom \cite{Kostelecky:2018yfa}. To date, $d = 3$ quark coefficients have been constrained using $K, D, B_d$, and $B_s$ meson oscillations~\cite{Kostelecky:1997mh, Kostelecky:1999bm, Kostelecky:2000ku, Isgur:2001yz, Nguyen:2001tg, Babusci:2013gda, Kostelecky:2001ff, Aubert:2007bp, Kostelecky:2010bk, Abazov:2015ana, Aaij:2016mos, Roberts:2017tmo, Schubert:2016xul}.  Relatively recently, techniques of chiral perturbation theory have established connections between minimal effective meson and baryon coefficients and parton coefficients, leading to several constraints on $d = 3, 4$ quark and $d = 4$ gluon coefficients using low-energy experiments and high-energy astrophysical sources \cite{Kamand:2016xhv,Kamand:2017bzl,Altschul:2019beo,Noordmans:2016pkr,Noordmans:2017kji}. Additional recent advances in the study of scalar fields have initiated investigations into the nonminimal meson sector, leading to constraints on effective $d=5$ coefficients for $K, D, B_d$, and $B_s$ mesons~\cite{Edwards:2018lsn, Edwards:2019lfb}. 

Constraints on hadron and quark coefficients from high-energy processes have typically come from the observed absence of cosmic-ray processes, including photon decay and vacuum Cherenkov radiation \cite{Gagnon:2004xh, Altschul:2006uw, Kostelecky:2013rta, Schreck:2017isa}. Constraints on heavy quarks are sparse in comparison to light quarks, however, a few bounds have been extracted from $e^+ e^-$ collisions at the Large Electron-Positron Collider (LEP), $t\bar t$ production at the Fermi National Accelerator Laboratory (Fermilab) and the LHC, radiative corrections, and potentially stable top hadrons from astrophysical sources \cite{Karpikov:2016qvq, Berger:2015yha, Abazov:2012iu, Carle:2019ouy, Satunin:2017wmk, Altschul:2020wkw}. Direct probes of minimal and nonminimal effects on initial-state partons can also be made through measurements of the scattering cross section in deep inelastic scattering and the DY process \cite{Kostelecky:2016pyx, Lunghi:2018uwj, Kostelecky:2019fse}. Collider processes such as these offer a complementary approach for constraining the space of parton coefficients, as the majority of existing quark-sector bounds from astrophysical sources are limited to the subset of isotropic (rotationally-invariant) coefficients. In contrast, experiments with Earth-based sources provide natural access to non-isotropic effects through sidereal modulations of scattering observables, enabling a broad search of the coefficient space.

In this work, we extend the recently developed formalism for addressing CPT- and Lorentz-violating effects in the collisions of high-energy hadrons at large momentum transfer \cite{Kostelecky:2019fse} to include spin-dependent effects. Precisely, we consider a set of dimensionless, CPT-even, parity-odd, and spin-dependent minimal SME coefficients affecting the propagation and interactions of quarks in the neutral-current DY process. It is demonstrated that the unpolarized DY process in the ultrarelativistic limit, through $\gamma/Z$ interference and pure $Z$ exchange, provides a natural avenue for accessing and constraining these coefficients. Note that effects of Lorentz violation on the $Z$ boson have also been studied in similar processes \cite{Fu:2016fmf, Colladay:2017zen, Michel:2019tti}, however, these considerations are outside the scope of this work. The leading-order cross section as a function of the dilepton invariant mass is derived.  Coupled with data on $Z$-boson production from the LHC, real and simulated constraints are placed on the coefficients for Lorentz violation for $u, d, s,$ and $c$ quarks.
\section{Theory}
\label{sec:theory}
\subsection{Interactions}
\label{sec:ints}
The dominant Lorentz-violating effects on quarks from the minimal SME are the subset of CPT-even interactions \cite{Colladay:1998fq, Colladay:2001wk}
\begin{align}
\label{model1}
\mathcal{L}^{\text{CPT}+} &\supset \tfrac{1}{2}i c_{Q_i}^{\mu\nu}\overline{Q}_i\gamma_\mu\overset{\text{\tiny$\leftrightarrow$}}{D_{\nu}}Q_i  + \tfrac{1}{2}i c_{U_i}^{\mu\nu}\overline{U}_i\gamma_\mu\overset{\text{\tiny$\leftrightarrow$}}{D_{\nu}}U_i 
 + \tfrac{1}{2}i c_{D_i}^{\mu\nu}\overline{D}_i\gamma_\mu\overset{\text{\tiny$\leftrightarrow$}}{D_{\nu}}D_i \;,
\end{align}
where $D^\nu$ is the conventional gauge-covariant derivative. The fields have the conventional definitions
\begin{align}
Q_i = \begin{pmatrix} u_i \\ d_i \end{pmatrix}_L, \quad U_i = \left(u_i\right)_R, \quad D_i = \left(d_i\right)_R, 
\end{align}
where $i = 1, 2, 3$ denotes flavors $u_i = (u, c, t)$ and $d_i = (d, s, b)$. The coefficients $c_{Q_i}^{\mu\nu}$, $c_{U_i}^{\mu\nu}$, and $c_{D_i}^{\mu\nu}$ transform as tensors under observer Lorentz transformations and are treated as perturbations with respect to conventional Lorentz-invariant effects. Hermiticity ensures real matrix elements, and energy-momentum conservation is preserved under the assumption the coefficients are spacetime constants. Field redefinitions render the antisymmetric parts of the coefficients unobservable at first order in Lorentz violation and the traces are Lorentz invariant \cite{Colladay:1998fq, Fittante:2012ua, Colladay:2002eh}. 
It is customary to define the coefficients
\begin{align}
\label{eq:quarkcoeffs}
c_{u_i}^{\mu\nu} &= (c_{Q_i}^{\mu\nu} + c_{U_i}^{\mu\nu})/2, \quad c_{d_i}^{\mu\nu} = (c_{Q_i}^{\mu\nu} + c_{D_i}^{\mu\nu})/2  \;, \nonumber\\
d_{u_i}^{\mu\nu} &= (c_{Q_i}^{\mu\nu} - c_{U_i}^{\mu\nu})/2, \quad d_{d_i}^{\mu\nu} = (c_{Q_i}^{\mu\nu} - c_{D_i}^{\mu\nu})/2  \;,
\end{align}
with the constraint $c_{u_i}^{\mu\nu} -c_{d_i}^{\mu\nu} = d_{d_i}^{\mu\nu} - d_{u_i}^{\mu\nu}$. 
\subsection{Partonic description of scattering}
The framework for the description of scattering in the presence of flavor-diagonal and spin-independent CPT- and Lorentz-violating operators of arbitrary mass dimension was recently developed in ref.~\cite{Kostelecky:2019fse}. We extend this approach by including minimal spin-dependent interactions, which requires additional attention. Given the perturbative approach, it is assumed throughout that effects at first order in Lorentz violation are considered. We begin with the modified kinetic Lagrange density stemming from eqs.~\eqref{model1}-\eqref{eq:quarkcoeffs} for a single massless fermion field $\psi$,
\begin{align}
&\mathcal{L} = \tfrac{1}{2}i\bar \psi\left(\eta^{\mu\nu} +  c^{\mu\nu} + d^{\mu\nu}\gamma_5\right)\gamma_\mu \overset{\text{\tiny$\leftrightarrow$}}{\partial_{\nu}}\psi.
\end{align}
Projecting out the chiral components of the resulting modified Dirac equation gives the following modified Weyl equations
\begin{align}
&i\widetilde{\slashed{\partial}}_L\psi_L = 0, \\
&i\widetilde{\slashed{\partial}}_R\psi_R= 0,
\end{align}
where $\psi_{L,R} = P_{L,R}\psi$ with the usual left and right projectors $P_{L} = (1-\gamma_5)/2$ and $P_{R} = (1+\gamma_5)/2$, respectively. The tilde variables are defined as
\begin{align}
&\widetilde{\partial}^\mu_L = (\eta^{\mu\nu} + c^{\mu\nu} + d^{\mu\nu})\partial_\nu \equiv \widetilde{\eta}^{\mu\nu}_L\partial_\nu,
\end{align}
and similarly for $\widetilde{\partial}^\mu_R$ with $d^{\mu\nu} \rightarrow -d^{\mu\nu}$. 
Multiplying by the associated Dirac operator and converting to momentum space yields the following dispersion relations
\begin{equation}
\label{moddispgen}
\widetilde{k}_{L,R}^2 = 0,
\end{equation}
with $\widetilde{k}^\mu_{L,R} \equiv \widetilde{\eta}^{\mu\nu}_{L,R}k_\nu$. Note that $\widetilde{k}_{L} \neq \widetilde{k}_{R}$ only for nonzero $d$ coefficients, implying species of definite handedness propagate according to dispersion relations related by a change of sign of the $d$ coefficients. This is transparent when considering the propagator may be written as 
\begin{fmffile}{diagram}
\begin{align}
\label{eq:quarkprop}
\begin{gathered}
\begin{fmfgraph*}(60,20)
\fmfleft{i1}
\fmfright{o1}
\fmf{fermion}{i1,v,o1}
\fmfdot{v}
\end{fmfgraph*}\vspace{-1.5mm}
\end{gathered} \quad&=  P_L\frac{i\slashed{\widetilde{k}}_L}{\widetilde{k}_L^2}+ P_R\frac{i\slashed{\widetilde{k}}_R}{\widetilde{k}_R^2}.
\end{align}
\end{fmffile}
The presence of nonzero $d$ coefficients splits the spin degeneracy leading to four distinct eigensolutions to the modified Dirac equation. In the ultrarelativistic limit this is clear upon decomposing the chiral spinors into the two-component form 
\begin{align}
\psi_L = \begin{pmatrix} 0 \\ \chi^{\pm} \end{pmatrix}, \quad \psi_R = \begin{pmatrix} \phi^{\pm} \\ 0  \end{pmatrix},
\end{align}
where $\pm$ denotes positive- and negative-energy solutions. After some calculation paralleling the conventional case, we find 
\begin{align}
\label{result}
&\hat{\widetilde{k}}_L\cdot\vec{\sigma}\chi^{\pm} = \mp \chi^{\pm},\\
&\hat{\widetilde{k}}_R\cdot\vec{\sigma}\phi^{\pm} = \pm \phi^{\pm},
\end{align}
where $\hat{\widetilde{k}}$ is a unit vector in the direction of $\widetilde{k}^j$.
In contrast to the Lorentz-invariant case, eigenstates of the chirality operator $\gamma_5$ are eigenstates of a modified helicity operator $\sim \hat{\widetilde{k}}\cdot \vec{\sigma}$. The conventional helicity operator $\sim \hat{k}\cdot \vec{\sigma}$ no longer commutes with Hamiltonian in the presence of nonzero $d$ coefficients, so helicity is no longer conserved. Instead, the eigenvalues of the operator corresponding to the spin projected along the momentum direction that satisfies the massless on-shell condition \eqref{moddispgen} are identified with states of definite chirality. In this limit, the $d$ coefficients act as a set of $c$ coefficients which change sign depending on the state's handedness. 

We wish to apply the above description to large momentum transfer processes. The original description of partons as nucleon constituents \cite{Feynman:1969ej}, as well as the field-theoretic description \cite{Collins:1989gx, Collins:2011zzd}, suggests that partons participating in deep-inelastic processes are reasonably approximated as massless physical particles with dispersion relation $k^2 \simeq 0$. In the present case, the on-shell condition is modified and given by eq.~\eqref{moddispgen}, $\widetilde{k}_{L,R}^2 = 0$. For an analogous partonic description, it is crucial to notice that observer Lorentz covariance is maintained in the SME and a covariant parametrization of the parton momentum in terms of its parent hadron is desired. For covariance to be maintained, in addition to the approximations of on-shell and massless partons, the unique choice of 
\begin{equation}
\widetilde{k}_{L,R}^\mu = \xi p^\mu
\label{eq:tildek}
\end{equation}
must be made, where $\xi \equiv \widetilde{k}_{L,R}^+/p^+$ is interpreted as the fraction of the parent hadron's momentum $p^\mu$ that is carried by the parton. The relevant procedure is therefore
to impose conditions \eqref{moddispgen}, \eqref{eq:tildek} as appropriate, from which the typical perturbative calculation of the scattering process parallels the conventional case. The dominant contribution to the resulting hadron cross section $\sigma_H$ for, e.g., the unpolarized DY process at leading order in electroweak interactions is found to take the schematic form
\begin{align}
\sigma_{H} \sim \sum_f\int d\xi d\xi' \hat{\sigma}_f(\xi, \xi')f_f(\xi)f_{\bar f}(\xi').
\end{align}
The perturbative (hard) partonic cross section $\hat{\sigma}_f(\xi, \xi')$ is integrated against the nonperturbative (soft) parton distribution functions (PDFs) for partons and antipartons. The PDFs may be expressed as hadron matrix elements of bilocal quark fields integrated along shifted light-cone directions:
\begin{align} 
& f_{fL,R}(\xi) = \int\fr{d\lambda,2\pi}e^{-i\xi p\cdot {n} \lambda}
\bra{p}\bar{\psi}_{fL,R}(\lambda \widetilde{n}^\mu_{fL,R})\frac{\slashed{n}}{2}\psi_{fL,R}(0)\ket{p},  \label{PDFs1} \\
&f_{\bar{f} L,R}(\xi) = -\int\fr{d\lambda,2\pi}e^{+i\xi p\cdot {n} \lambda}
\bra{p}\bar{\psi}_{fL,R}(\lambda \widetilde{n}^\mu_{fL,R})\frac{\slashed{n}}{2}\psi_{fL,R}(0)\ket{p}. \label{PDFs2}
\end{align}
These expressions along with their potential dependence on the coefficients for Lorentz violation will be derived in explicit calculations in Sec.~\ref{sec:Zexch}.

In a particular observer frame, the PDFs describe the probability for finding a left- or right-handed parton with momentum fraction $\xi$ such that $\widetilde{k}_{L,R} = \xi p$. However, given that the PDFs are generically functions of $\xi, Q^2$ and potentially other Lorentz scalars, the functions themselves are Lorentz invariant in the conventional case and Lorentz observer invariant (but \textit{not} particle invariant) in the Lorentz-violating case. The same considerations apply to the partonic cross sections $\hat{\sigma}_f(\xi, \xi')$. These former features can be understood in the context of local and Lorentz observer-invariant operator products \cite{Kostelecky:2016pyx, Kostelecky:2019fse},
\begin{align}
\mathcal{O}^{\mu_1\cdots\mu_n}_{fL,R} 
= \bar{\psi}_{fL,R}\gamma^{\{\mu_1}
(i\tilde{D}_{L,R}^{\mu_2})(i\tilde{D}_{L,R}^{\mu_3})\ldots
(i\tilde{D}_{L,R}^{\mu_n\}})\psi_f-\text{traces} .
\label{eq:Optwist2}
\end{align}
Taking the hadron matrix elements and contracting with external light-cone vectors shows that the matrix-element coefficients are identified with the moments of the PDFs given by eqs.~\eqref{PDFs1}-\eqref{PDFs2}. This connection is crucial for determining how the PDFs potentially depend on the coefficients for Lorentz violation.  Establishing whether these features hold in the presence of radiative corrections and DGLAP evolution is worthy of further investigation but is outside the scope of this work. 

\section{The Drell-Yan process}
As an explicit example of the formalism of the previous section, we study the unpolarized neutral-current DY process $p_1 + p_2 \rightarrow \gamma/Z \rightarrow l_1+ l_2 + X$ at leading order, including the Lorentz-violating effects on the quarks as described in eq.~\eqref{model1}. At this level the process is initiated by the annihilation of a quark-antiquark pair of the same flavor, each residing in either of the two initial-state hadrons, producing a lepton pair $l_1 + l_2$ and an unmeasured hadronic final state $X$. 
In the following section, we calculate the cross section for this process. Using LHC events near and below the $Z$-boson pole, first constraints are placed on several coefficients that produce time-independent shifts to the conventional Lorentz-invariant result. For coefficients that induce time-dependent effects, simulations are performed to extract estimated constraints by binning the cross section as a function of sidereal time.

\subsection{Setup}
In the limit of massless quarks, it is advantageous to work in terms of chiral fields $\psi_{L,R}$. We introduce chiral SME coefficients and shifted Minkowski metrics
\begin{align}
&c_{fL}^{\mu\nu} \equiv c_f^{\mu\nu} + d_f^{\mu\nu}, \quad c_{fR}^{\mu\nu} \equiv c_f^{\mu\nu} - d_f^{\mu\nu}, \quad \widetilde{\eta}_{fL,R}^{\mu\nu} \equiv \eta^{\mu\nu} + c_{fL,R}^{\mu\nu},
\label{eq:coefficients}
\end{align}
for quark flavor $f$. Note that $c_{uL}^{\mu\nu} = c_{dL}^{\mu\nu} = c_{Q_1}^{\mu\nu}$, $c_{uR}^{\mu\nu} = c_{U_1}^{\mu\nu}$, and $c_{dR}^{\mu\nu} = c_{D_1}^{\mu\nu}$ (and similarly for the second and third generations).
The subset of interactions and notation introduced thus far leads to a model Lagrange density expressed in terms of quantities below the electroweak scale
\begin{align}
\label{model2}
\mathcal{L} = \sum_f\tfrac{1}{2}i\bar\psi_{fL}\widetilde{\eta}^{\nu\mu}_{fL}\gamma_\nu\left(\overset{\text{\tiny$\leftrightarrow$}}{\partial_{\mu}} - 2i(e e_fA_\mu + g_Zg_{fL}Z_\mu)\right)\psi_{fL} + (L\rightarrow R),
\end{align}
where the massless and massive gauge boson fields are $A_\mu$ and $Z_\mu$, respectively. The quark charges are $e_f$ and the couplings of the $Z$ boson in terms of the weak mixing angle $\theta_W$ are $g_Z = e/\sin\theta_W\cos\theta_W$, $g_{fL} = I_W^{(3)} - e_f \sin^2\theta_W$, and $g_{fR} = -e_f\sin^2\theta_W$.

At energies much greater than the hadron masses the differential cross section reads 
\begin{align}
\label{crosssec}
&d\sigma = \frac{1}{2s}\frac{d^3 l_1}{(2\pi)^3l_1^0}\frac{d^3 l_2}{(2\pi)^32l_2^0}\sum_X|\braket{l_1, l_2, X|\widehat{T}|p_1, s_1, p_2, s_2}|^2,
\end{align}
where $|\braket{l_1, l_2, X|\widehat{T}|p_1, s_1, p_2, s_2}|^2 = (2\pi)^4\delta^4(p_1 + p_2 - l_1 - l_2 - p_X)|\mathcal{M}_\gamma + \mathcal{M}_Z|^2$ \cite{Kostelecky:2019fse}. The amplitude has the form 
\begin{align}
\mathcal{M} =  \bar u(l_1)g\gamma^\mu(g_LP_L + g_RP_R)\nu(l_2)D_{\mu\nu}(q,M)\bra{X}j^\nu(0)\ket{p_1, s_1, p_2, s_2},  \label{genamp}
\end{align}
where $g$ is the electroweak vertex coupling, $g_L, g_R$ are the left and right lepton couplings, $D_{\mu\nu}(q,M)$ is the intermediate propagator, and $\bra{X}j^\nu(0)\ket{p_1, s_1, p_2, s_2}$ the hadron current matrix element. The spin indices of the leptons are suppressed as the polarizations of the final leptons are not measured. Combining the momentum-conserving delta function with the sum over intermediate hadron states and the hadron matrix elements yields the tensor 
\begin{align}
\label{hadronictensor}
& W^{\mu\nu} =  \int d^4x e^{-iq\cdot x}\bra{p_1, s_1, p_2, s_2}j^{\dagger \mu}(x)j^\nu(0)\ket{p_1, s_1, p_2, s_2}.
\end{align}
Calculating the unpolarized cross section requires an average over initial hadron spins and sum over final lepton spins for each of the four terms $|\mathcal{M}_\gamma + \mathcal{M}_Z|^2 = |\mathcal{M}_\gamma|^2 + \mathcal{M}^*_\gamma\mathcal{M}_Z + \mathcal{M}_\gamma\mathcal{M}^*_Z + |\mathcal{M}_Z|^2$. The pure electromagnetic contribution has been calculated and is given by eq.~(4.18) in ref.~\cite{Kostelecky:2019fse}. An interesting feature of this contribution is that the $d$ coefficients produce no observable effect---this occurs because they generate totally antisymmetric contributions to the hadron tensor, whereas the lepton tensor for this contribution is symmetric. One might think that the absence of contributions from the $d$ coefficients in the hard process may not necessarily imply nonperturbative contributions from the PDFs will be absent. However, the equality of left- and right-handed couplings in electromagnetic interactions implies that potential contributions $\sim \pm d^{pp}$ would be canceled upon expanding the PDFs to first order in Lorentz violation. Therefore, we are primarily concerned with effects stemming from interference $ \mathcal{M}^*_\gamma\mathcal{M}_Z + \mathcal{M}_\gamma\mathcal{M}^*_Z$ and pure $Z$ exchange $|\mathcal{M}_Z|^2$.
\subsection{Illustration of the pure $Z$-exchange contribution}
\label{sec:Zexch}
It is illustrative to show some of the required calculations in detail. For this purpose, we outline the $|\mathcal{M}_Z|^2$ contribution. The reader is encouraged to consult ref.~\cite{Kostelecky:2019fse} for additional details and discussion. 

We begin by expanding the current $j^\mu_Z$ in terms of chiral components:
\begin{align}
j^\mu_Z &= g_Z\bar\psi_f\Gamma_f^\mu(g_{fL}P_L + g_{fR}P_R)\psi_f  \\
&= g_Z\left[g_{fL}\bar\psi_{fL}\Gamma^\mu_{fL} \psi_{fL} + g_{fR}\bar\psi_{fR}\Gamma^\mu_{fR}\psi_{fR}\right],
\end{align}
where $\Gamma^\mu_{fL,R} \equiv (\eta^{\nu\mu} + c_{fL,R}^{\nu\mu})\gamma_\nu$. The relevant current product in eq.~\eqref{hadronictensor} admits several Dirac structures; however, a single term dominates in the leading-twist approximation for unpolarized scattering \cite{Jaffe:1997vlv, Jaffe:1991ra}, giving
\begin{align}
\label{jzcurrentprod}
j_Z^{\dagger\mu}(x)j_Z^\nu(0) \simeq -\frac{g_Z^2}{12}&\left(g_{fL}^2\text{Tr}\left[P_L\Gamma^\mu_{fL}\gamma^\rho P_L \Gamma^\nu_{fL}\gamma^\sigma\right] \right. \nonumber\\
&\left. \times \left[\bar\psi_{fL}(0)\gamma_\rho\psi_{fL}(x)\right]\left[\bar\psi_{fL}(x)\gamma_\sigma\psi_{fL}(0)\right]  + (L\rightarrow R)\right).
\end{align}
From here it is useful to choose an observer frame. We choose the hadrons' center-of-mass (CM) frame and parametrize the momenta as $p_1 = p_1^+ \bar{n}$, $p_2 = p_2^- n$, where $\bar{n}, n$ are two lightlike vectors $\bar n^\mu = (1,0,0,+1)/\sqrt{2}, n^\mu = (1,0,0,-1)/\sqrt{2}$.
This choice will yield large $+ (-)$ components for the $\ket{p_1} (\ket{p_2})$ matrix elements. Identifying the combination of gamma matrices that project out the large components of the matrix elements, arranging the bilinear operators to preserve colorless matrix elements, and employing the Sudakov decomposition gives the dominant contribution to the hadron tensor for a given flavor $f$ as
\begin{align}
\label{Wpreparton}
W_f^{\mu\nu} \simeq  -&\frac{g_Z^2g_{fL}^2}{4N_cp_1^+ p_2^-} \text{Tr}\left[P_L\Gamma_{fL}^\mu\slashed{p}_1P_L\Gamma^\nu_{fL}\slashed{p}_2\right] \nonumber\\*
& \times\int d^4xe^{-iq\cdot x}\bra{p_1}\bar{\psi}_{fL}(x)\gamma^+\psi_{fL}(0)\ket{p_1} \bra{p_2}\bar{\psi}_{fL}(0)\gamma^-\psi_{fL}(x)\ket{p_2} + (L\rightarrow R),
\end{align}
where $N_c = 3$ for the $SU(3)_c$ gauge group. With the appropriate dispersion relation \eqref{moddispgen} taken into account, each term may be independently factorized according to the known procedure \cite{Kostelecky:2019fse}. After some calculation the hadron tensor takes the form
\begin{align}
\label{hadronictensorf}
W_f^{\mu\nu} = \frac{\widetilde{s}}{2}\int d\xi_1d\xi_2 \left\{H_{fL}^{\mu\nu}(\xi_1,\xi_2)
\left[f_{fL}(\xi_1)f_{\bar{f}L}(\xi_2) 
+ f_{fL}(\xi_2)f_{\bar{f}L}(\xi_1)\right] + (L\rightarrow R)\right\}, 
\end{align}
where hard scattering coefficient functions are
\begin{align}
\label{hardscattZ}
H^{\mu\nu}_{fL,R}(\xi_1,\xi_2) = 
\fr{8g_Z^2g_{fL,R}^2,N_c\widetilde{s}}&\text{Tr}\left[P_L\Gamma^\mu_{fL,R}\fr{\xi_1 \slashed{p}_1,2}P_L\Gamma^\nu_{fL,R}\fr{\xi_2\slashed{p}_2,2}\right]
\nonumber\\
&\times 
(2\pi)^4\delta^4\left(q^\mu + \xi_1( c_{fL,R}^{\mu p_1} -p_1^\mu )+ \xi_2 (c_{fL,R}^{\mu p_2}-p_2^\mu)\right),
\end{align}
with $\widetilde{s}\equiv 2\xi_1\xi_2p_1\cdot p_2$.
Note that eqs.~\eqref{Wpreparton}-\eqref{hardscattZ} are valid in any observer frame by Lorentz observer covariance.
The hadron tensor represents two possibilities: one where the incident parton pair are eigenstates of definite left chirality, and the other definite right chirality. This can be seen by inspecting the hard scattering coefficient \eqref{hardscattZ}---the trace terms are exactly what one finds for a left- or right-handed parton pair annihilating with an on-shell momentum parameterization $\widetilde{k}_{L,R} = \xi p$. The PDFs in eq.~\eqref{hadronictensorf} are given by eqs.~\eqref{PDFs1}-\eqref{PDFs2}, which we repeat here for completeness:
\begin{align}
& f_{fL,R}(\xi, c_{fL,R}^{pp}/\Lambda^2_{\text{QCD}}) = \int\fr{d\lambda,2\pi}e^{-i\xi p\cdot {n} \lambda}
\bra{p}\bar{\psi}_{fL,R}(\lambda \widetilde{n}^\mu_{fL,R})\frac{\slashed{n}}{2}\psi_{fL,R}(0)\ket{p}, \label{SMEPDFprodf}\\
&f_{\bar{f}L,R}(\xi, c_{fL,R}^{pp}/\Lambda^2_{\text{QCD}})) = -\int\fr{d\lambda,2\pi}e^{+i\xi p\cdot {n} \lambda}
\bra{p}\bar{\psi}_{fL,R}(\lambda \widetilde{n}^\mu_{fL,R})\frac{\slashed{n}}{2}\psi_{fL,R}(0)\ket{p}.  \label{SMEPDFprodbarf}
\end{align}
In the presence of Lorentz-violating effects, the PDFs have the potential to depend on additional dimensionless scalar quantities that are functions of the external hadron momentum. Here, the explicit dependence on $c^{\mu\nu}_{fL,R}$ in the field $\bar{\psi}_{fL,R}(\lambda \widetilde{n}^\mu_{fL,R})$ translates, via considerations of the product of local operators and the reparametrization invariance of the PDFs, to an implicit dependence of $c_{fL,R}^{pp}/\Lambda^2_{\text{QCD}}$.
\subsection{Cross section}
Calculation of the remaining interference terms $\mathcal{M}^*_\gamma\mathcal{M}_Z + \mathcal{M}_\gamma\mathcal{M}^*_Z$ follows similarly as the calculation of $|\mathcal{M}_Z|^2$ from the previous section, differing only in the relevant currents and couplings. To construct the differential distribution $d\sigma/dQ^2$, where $q^2 \equiv Q^2 > 0$, it is simplest to evaluate eq.~\eqref{crosssec} in the dilepton CM frame. Contracting the lepton and hadron tensors and integrating over the solid angle of the final-state leptons produces a Lorentz observer-invariant function of the external hadron momenta. The final dilepton distribution is found to be
\begin{align}
\label{DYsigmadQ2}
&\frac{d\sigma}{dQ^2} =  \frac{4\pi\alpha^2}{3 N_c}\sum_f \left[\frac{e_f^2}{2Q^4} +\frac{1-m_Z^2/Q^2}{(Q^2 -m_Z^2)^2 + m_Z^2\Gamma_Z^2} \frac{1-4\sin^2\theta_W}{4\sin^2\theta_W\cos^2\theta_W}e_fg_{fL}  \right. \nonumber\\
&\left. + \frac{1}{(Q^2 -m_Z^2)^2 + m_Z^2\Gamma_Z^2}\frac{1+(1-4\sin^2\theta_W)^2}{32\sin^4\theta_W\cos^4\theta_W}g_{fL}^2\right]
\int_{\tau}^1 dx\frac{\tau}{x}\hat{\sigma}'_{f}\left(x,\tau/x,c_{fL}^{\mu\nu}\right) + (L\rightarrow R),
\end{align}
where $\tau \equiv Q^2/s$ is the scaling variable, $m_Z, \Gamma_Z$ are the mass and width of the $Z$ boson, and 
\begin{align}
\label{sigmaprime}
\hat{\sigma}'_{f}\left(x,\tau/x, c_{fL}^{\mu\nu}\right) &\equiv \left( 1+ \frac{2}{s}c_{fL}^{\mu\nu}(1+x^2/\tau)(p_{1\mu}p_{1\nu}+ p_{1\mu}p_{2\nu}+ (p_1\leftrightarrow p_2))\right)  f_{f\text{S}}(x,\tau/x) \nonumber\\
&\quad +  \frac{2}{s}c_{fL}^{\mu\nu}\left(xp_{1\mu}p_{1\nu} +\frac{\tau}{x}p_{1\mu}p_{2\nu} + (p_1\leftrightarrow p_2) \right) f'_{f\text{S}}(x,\tau/x), 
\end{align}
where the flavor-symmetric PDF products are defined as
\begin{align}
&f_{f\text{S}}(x,\tau/x) \equiv f_{f}(x)f_{\bar f}(\tau/x) +  f_{f}(\tau/x)f_{\bar f}(x), \label{PDFprodf}\\
&f'_{f\text{S}}(x,\tau/x) \equiv f_{f}(x)f'_{\bar f}(\tau/x) +  f'_{f}(\tau/x)f_{\bar f}(x). \label{PDFprodbarf}
\end{align}
The first contribution in eq.~\eqref{DYsigmadQ2} is due to pure electromagnetic exchange $|\mathcal{M}_\gamma|^2$ and matches eq.~(4.18) in ref.~\cite{Kostelecky:2019fse} when evaluated in the collider frame and accounting for the sum over left and right coefficients. The second and third terms are due to interference $\mathcal{M}^*_\gamma\mathcal{M}_Z + \mathcal{M}_\gamma\mathcal{M}^*_Z$ and pure $Z$ exchange $|\mathcal{M}_Z|^2$, respectively.
Notice that eq.~\eqref{sigmaprime} differs from the conventional result $\tau f_{f\text{S}}(x,\tau/x)$ by terms proportional to the coefficients for Lorentz violation and reduces to the Lorentz-invariant result in the limit of vanishing coefficients \cite{Quigg:2013ufa, Brock:1993sz}. The correct symmetrization properties of the coefficients, as dictated from field redefinitions, are also displayed. 

Parity invariance of QCD connects the PDF products \eqref{PDFprodf}-\eqref{PDFprodbarf} to those of eqs.~\eqref{SMEPDFprodf}-\eqref{SMEPDFprodbarf}. However, as the $d$ coefficients are odd under parity, additional corrections of $\mathcal{O}(c_{f}^{pp})$ to the electromagnetic contribution and $\mathcal{O}(c_{fL}^{pp}, c_{fR}^{pp})$ to the interference and $Z$ contributions may stem from the expansion of the PDFs to first order in Lorentz violation. Given the inherently nonperturbative albeit unknown origin of these additional contributions \cite{Vieira:2019nxg}, we have suppressed them in the final result \eqref{DYsigmadQ2}. While these effects constitute an interesting open issue, they have little bearing on extracting bounds on the coefficients for Lorentz violation and are neglected in the analysis that follows. 

%
The cross section \eqref{DYsigmadQ2} enables a comparison with the entire $Q^2$ spectrum measured at LHC. In ref.~\cite{Kostelecky:2019fse} it was shown that the photon contribution is dominated by measurements at low $Q^2$~\cite{Sirunyan:2018owv}. We revisit the impact of measurements at low $Q^2$ and present estimated bounds using total cross-section measurements on the $Z$ pole by the CMS collaboration~\cite{Sirunyan:2019bzr}.
The main difference between photon and $Z$ interactions is the presence of parity violation which, in turn, introduces dependence on the parity-odd $d$ coefficients defined in eq.~\eqref{eq:quarkcoeffs}. At low $Q^2$, eq.~\eqref{eq:quarkcoeffs} includes the interference between diagrams mediated by the exchange of a photon or $Z$ boson. Thus, we expect some sensitivity to the $d$ coefficients in addition to the parity-even $c$ coefficients away from the $Z$ pole. On the $Z$ pole, parity-violating effects are maximal, and we expect a strong sensitivity to the $d$ coefficients. 

\begin{figure}[ht]
\centering
\includegraphics[width=0.69\linewidth]{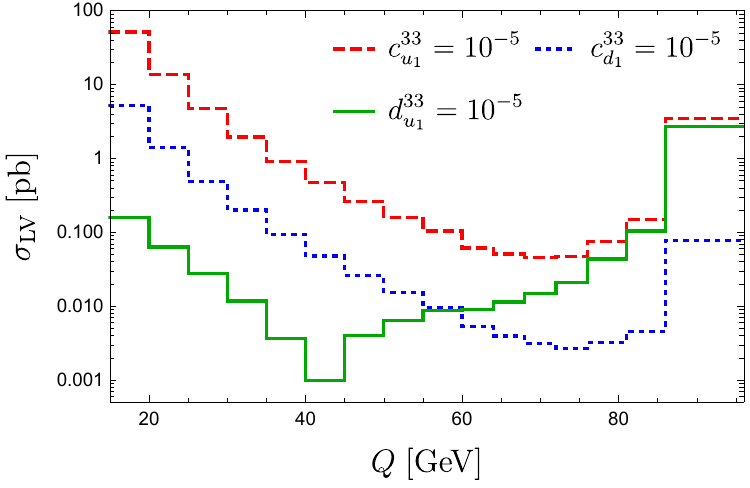}
\caption{An illustration of the Lorentz-violating contributions $\sigma_{\text{LV}}$ from eq.~(\ref{DYsigmadQ2}) in the collider frame for select coefficients fixed to magnitudes of $10^{-5}$.}
\label{figure1} 
\end{figure}

To highlight these issues, we present the results using two bases: the ($c_{U_i}$, $c_{D_i}$, $c_{Q_i}$) basis introduced in eq.~\eqref{model1}, and the ($c_{u_i}$, $c_{d_i}$, $d_{u_i}$) basis defined in eq.~\eqref{eq:quarkcoeffs}, with $d_{d_i} = d_{u_i} + c_{u_i} - c_{d_i}$. The former choice is the most natural in situations in which parity is maximally broken. In this case the bounds on $c_{U_i}$, $c_{D_i}$, and $c_{Q_i}$  are expected to be mostly independent. The latter is more useful to disentangle correlations between the expected bounds for situations in which the $d_{u_i,d_i}$ coefficients are poorly constrained (as in DY at low $Q^2$).  These features can be understood by inspecting the Lorentz-violating part of eq.~(\ref{DYsigmadQ2}) in the collider frame, for which it can be shown that only the coefficients $c_{fL,R}^{33}$ and  $c_{fL,R}^{00}$ appear. In Fig.~\ref{figure1}, a comparison is made for combinations of the former coefficients for several values of $Q \in [17.5, 90]$ GeV. It is observed that all coefficients have decreasing sensitivity from low $Q$ until approaching the $Z$ pole, in agreement with the prior results of ref.~\cite{Kostelecky:2019fse}. Near the $Z$ pole, the $d$-type coefficients produce the greatest signal, as expected from prior considerations of parity violation. 

We remind the reader that the formalism presented in Sec.~\ref{sec:theory} applies in the limit of massless quarks, implying that the validity of any derived constraints is contingent upon the scale of the momentum transfer. As we consider data for $Q \lesssim 100$ GeV, reasonable limits can only be extracted for the first two generations---the hard scattering is thus sensitive to $u$, $d$, $s$, and $c$ partons and bounds in the following sections are extracted for these flavors. 

\subsection{Time-independent bounds}
 We follow the convention of reporting constraints on coefficient combinations as they appear in the Sun-centered frame with coordinates $\widehat{T}, \widehat{X}, \widehat{Y}, \widehat{Z}$ \cite{Kostelecky:2002hh,Bluhm:2001rw,Bluhm:2003un}.
The Sun-centered frame and laboratory (collider) frame are related by a rotation to an excellent approximation, yielding the relations $c_{fL,R}^{00} = c_{fL,R}^{TT}$ and 
\begin{align}
c_{fL,R}^{33} =& 
\frac{1}{2}(c_{fL,R}^{XX} + c_{fL,R}^{YY})\left(\cos^2\chi\sin^2\psi + \cos^2\psi\right) +  c_{fL,R}^{ZZ}\sin^2\chi\sin^2\psi
\nonumber\\
& - 2c_{fL,R}^{XZ}\sin\chi\sin\psi
\left[\cos\chi\sin\psi\cos(\Omega_{\oplus}T_{\oplus} )
+ \cos\psi\sin(\Omega_{\oplus}T_{\oplus})\right] 
\nonumber\\
& - 2c_{fL,R}^{YZ}\sin\chi\sin\psi
\left[\cos\chi\sin\psi\sin(\Omega_{\oplus}T_{\oplus}) 
- \cos\psi\cos(\Omega_{\oplus}T_{\oplus})\right] \nonumber\\
& + c_{fL,R}^{XY}\left[(\cos^2\chi\sin^2\psi - \cos^2\psi)\sin(2\Omega_\oplus T_\oplus) - \cos\chi\sin(2\psi)\cos(2\Omega_\oplus T_\oplus)\right] \nonumber \\
& +\frac{1}{2}(c_{fL,R}^{XX} - c_{fL,R}^{YY})\left[(\cos^2\chi\sin^2\psi - \cos^2\psi)\cos(2\Omega_\oplus T_\oplus) \right. \nonumber\\
 & \left.+ \cos\chi\sin(2\psi)\sin(2\Omega_\oplus T_\oplus)\right].
\label{eq:c33rot}
\end{align}
Notice that the coefficients with indices of the form $TT, XX + YY, ZZ$ are associated with time-independent effects, whereas those of the forms $XZ, YZ$ and $XY, XX-YY$ are associated with first and second harmonics of the Earth's sidereal frequency $\Omega_\oplus$ and local sidereal time $T_\oplus$, respectively.

The time-independent bounds are given in table~\ref{tab:timeindep} and depend on the difference between the measured cross section and the SM prediction. Moreover, theory uncertainties are taken into account for the latter by combining with statistical and systematic errors in quadrature. We employ SM predictions calculated 
using the FEWZ package~\cite{Melnikov:2006di, Gavin:2010az, Gavin:2012sy, Li:2012wna} at NNLO accuracy using the NNPDF 3.1~\cite{Ball:2017nwa} NNLO PDF set.

\begin{table}[t]
\centering
\begin{tabular}{|c||c|c|}\hline
coefficient & $[\frac{d\sigma}{dQ}]_{Q=17.5 \; {\rm GeV}}$ & $[\frac{d\sigma}{dQ}]_{Q=m_Z}$ \\ \hline
$|c_{u_1}^{TT}|$ & $1.2\times 10^{-4}$ & $8.9\times 10^{-4}$ \\ 
$|c_{u_1}^{XX}+c_{u_1}^{YY}|$ & $1.4\times 10^{-4}$ & $8.5\times 10^{-4}$ \\ 
$|c_{u_1}^{ZZ}|$ & $4.1\times 10^{-2}$ & $2.4\times 10^{-1}$ \\ \hline
$|c_{d_1}^{TT}|$ & $1.2\times 10^{-3}$ & $4.0\times 10^{-2}$ \\ 
$|c_{d_1}^{XX}+c_{d_1}^{YY}|$ & $1.4\times 10^{-3}$ & $3.8\times 10^{-2}$ \\ 
$|c_{d_1}^{ZZ}|$ & $4.0\times 10^{-1}$ & $-$ \\ \hline
$|d_{u_1}^{TT}|$ & $4.0\times 10^{-2}$ & $1.1\times 10^{-3}$ \\ 
$|d_{u_1}^{XX}+d_{u_1}^{YY}|$ & $4.6\times 10^{-2}$ & $1.1\times 10^{-3}$ \\ 
$|d_{u_1}^{ZZ}|$ & $-$ & $3.1\times 10^{-1}$ \\ \hline\hline
$|c_{u_2}^{TT}|$ & $7.4\times 10^{-3}$ & $1.8\times 10^{-2}$ \\ 
$|c_{u_2}^{XX}+c_{u_2}^{YY}|$ & $8.2\times 10^{-3}$ & $1.6\times 10^{-2}$ \\ 
$|c_{u_2}^{ZZ}|$ & $-$ & $-$ \\ \hline
$|c_{d_2}^{TT}|$ & $1.2\times 10^{-2}$ & $3.5\times 10^{-1}$ \\ 
$|c_{d_2}^{XX}+c_{d_2}^{YY}|$ & $1.3\times 10^{-2}$ & $3.3\times 10^{-1}$ \\ 
$|c_{d_2}^{ZZ}|$ & $-$ & $-$ \\ \hline
$|d_{u_2}^{TT}|$ & $-$ & $1.9\times 10^{-2}$ \\ 
$|d_{u_2}^{XX}+d_{u_2}^{YY}|$ & $-$ & $1.8\times 10^{-2}$ \\ 
$|d_{u_2}^{ZZ}|$ & $-$ & $-$ \\ \hline\hline
\end{tabular}
\caption{Bounds on the time-independent coefficients in the Sun-centered frame and in the  $(c_{u_i} , c_{d_i}, d_{u_i})$ basis for $Q=m_Z$ and $Q = 17.5$\;GeV. The bound on $d_{u_2}^{ZZ}$ has no sensitivity at $Q=m_Z$ and has been excluded. 
\label{tab:timeindep}}
\end{table}
\noindent The bounds listed in table~\ref{tab:timeindep} support the general trend shown in Fig.~\ref{figure1} that $c$ ($d$)-type coefficients are more sensitive to lower (higher) values of $Q$. A number of these bounds represent first constraints on $u, d, s$, and $c$ quarks, in particular those with the indices $XX + YY$ and $ZZ$; however, constraints on the isotropic $TT$ coefficients are not competitive with existing constraints \cite{Gagnon:2004xh, Kostelecky:2013rta, Schreck:2017isa}. 
\subsection{Time-dependent bounds}

\begin{table}[t]
\centering
\begin{tabular}{|c||c|c|c|}\hline
\multirow{2}{*}{coefficient} & 
\multicolumn{3}{c|}{
$[\frac{d\sigma}{dQ}]_{Q=17.5 \; {\rm GeV}}$}  \\ \cline{2-4}
 & $\delta_{\rm th}$ & $\delta_{\rm th}$, $\delta_{\rm lumi}$ &
  $\delta_{\rm th}$, $\delta_{\rm lumi}$, $\delta_{\rm sel}$ \\ \hline
$|c_{U_1}^{XY}|$ & $5.2\times 10^{-5}$ & $4.7\times 10^{-5}$ & $2.1\times 10^{-5}$ \\ 
$|c_{U_1}^{XZ}|$ & $1.4\times 10^{-4}$ & $1.3\times 10^{-4}$ & $5.9\times 10^{-5}$ \\ 
$|c_{U_1}^{YZ}|$ & $1.4\times 10^{-4}$ & $1.3\times 10^{-4}$ & $5.9\times 10^{-5}$ \\ 
$|c_{U_1}^{XX}-c_{U_1}^{YY}|$ & $2.8\times 10^{-4}$ & $2.5\times 10^{-4}$ & $1.1\times 10^{-4}$ \\ \hline
$|c_{D_1}^{XY}|$ & $5.1\times 10^{-4}$ & $4.6\times 10^{-4}$ & $2.1\times 10^{-4}$ \\ 
$|c_{D_1}^{XZ}|$ & $1.4\times 10^{-3}$ & $1.3\times 10^{-3}$ & $5.8\times 10^{-4}$ \\ 
$|c_{D_1}^{YZ}|$ & $1.4\times 10^{-3}$ & $1.3\times 10^{-3}$ & $5.8\times 10^{-4}$ \\ 
$|c_{D_1}^{XX}-c_{D_1}^{YY}|$ & $2.7\times 10^{-3}$ & $2.4\times 10^{-3}$ & $1.1\times 10^{-3}$ \\ \hline
$|c_{Q_1}^{XY}|$ & $4.7\times 10^{-5}$ & $4.3\times 10^{-5}$ & $1.9\times 10^{-5}$ \\ 
$|c_{Q_1}^{XZ}|$ & $1.3\times 10^{-4}$ & $1.2\times 10^{-4}$ & $5.4\times 10^{-5}$ \\ 
$|c_{Q_1}^{YZ}|$ & $1.3\times 10^{-4}$ & $1.2\times 10^{-4}$ & $5.4\times 10^{-5}$ \\ 
$|c_{Q_1}^{XX}-c_{Q_1}^{YY}|$ & $2.5\times 10^{-4}$ & $2.3\times 10^{-4}$ & $1.0\times 10^{-4}$ \\ \hline\hline
$|c_{U_2}^{XY}|$ & $3.0\times 10^{-3}$ & $2.7\times 10^{-3}$ & $1.2\times 10^{-3}$ \\ 
$|c_{U_2}^{XZ}|$ & $8.3\times 10^{-3}$ & $7.5\times 10^{-3}$ & $3.4\times 10^{-3}$ \\ 
$|c_{U_2}^{YZ}|$ & $8.4\times 10^{-3}$ & $7.6\times 10^{-3}$ & $3.4\times 10^{-3}$ \\ 
$|c_{U_2}^{XX}-c_{U_2}^{YY}|$ & $1.6\times 10^{-2}$ & $1.4\times 10^{-2}$ & $6.5\times 10^{-3}$ \\ \hline
$|c_{D_2}^{XY}|$ & $4.8\times 10^{-3}$ & $4.3\times 10^{-3}$ & $2.0\times 10^{-3}$ \\ 
$|c_{D_2}^{XZ}|$ & $1.3\times 10^{-2}$ & $1.2\times 10^{-2}$ & $5.5\times 10^{-3}$ \\ 
$|c_{D_2}^{YZ}|$ & $1.3\times 10^{-2}$ & $1.2\times 10^{-2}$ & $5.4\times 10^{-3}$ \\ 
$|c_{D_2}^{XX}-c_{D_2}^{YY}|$ & $2.5\times 10^{-2}$ & $2.3\times 10^{-2}$ & $1.0\times 10^{-2}$ \\ \hline
$|c_{Q_2}^{XY}|$ & $1.9\times 10^{-3}$ & $1.7\times 10^{-3}$ & $7.6\times 10^{-4}$ \\ 
$|c_{Q_2}^{XZ}|$ & $5.1\times 10^{-3}$ & $4.7\times 10^{-3}$ & $2.1\times 10^{-3}$ \\ 
$|c_{Q_2}^{YZ}|$ & $5.2\times 10^{-3}$ & $4.7\times 10^{-3}$ & $2.1\times 10^{-3}$ \\ 
$|c_{Q_2}^{XX}-c_{Q_2}^{YY}|$ & $9.9\times 10^{-3}$ & $8.9\times 10^{-3}$ & $4.0\times 10^{-3}$ \\ \hline\hline
\end{tabular}
\caption{Expected best constraints on the time-dependent coefficients in the ($c_{U_i}$, $c_{D_i}$, $c_{Q_i}$) basis for $Q=17.5\; {\rm GeV}$. The part of the experimental uncertainties that are assumed to be 100\% correlated between binned data is indicated in the column label.
\label{tab:Q171}}
\end{table}

\begin{table}[t]
\centering
\begin{tabular}{|c||c|c|c|}\hline
\multirow{2}{*}{coefficient} & 
\multicolumn{3}{c|}{
$[\frac{d\sigma}{dQ}]_{Q=17.5 \; {\rm GeV}}$}  \\ \cline{2-4}
 & $\delta_{\rm th}$ & $\delta_{\rm th}$, $\delta_{\rm lumi}$ &
  $\delta_{\rm th}$, $\delta_{\rm lumi}$, $\delta_{\rm sel}$ \\ \hline
$|c_{u_1}^{XY}|$ & $2.6\times 10^{-5}$ & $2.3\times 10^{-5}$ & $1.1\times 10^{-5}$ \\ 
$|c_{u_1}^{XZ}|$ & $7.4\times 10^{-5}$ & $6.4\times 10^{-5}$ & $3.0\times 10^{-5}$ \\ 
$|c_{u_1}^{YZ}|$ & $7.4\times 10^{-5}$ & $6.5\times 10^{-5}$ & $3.0\times 10^{-5}$ \\ 
$|c_{u_1}^{XX}-c_{u_1}^{YY}|$ & $1.4\times 10^{-4}$ & $1.2\times 10^{-4}$ & $5.6\times 10^{-5}$ \\ \hline
$|c_{d_1}^{XY}|$ & $2.5\times 10^{-4}$ & $2.3\times 10^{-4}$ & $1.0\times 10^{-4}$ \\ 
$|c_{d_1}^{XZ}|$ & $7.2\times 10^{-4}$ & $6.3\times 10^{-4}$ & $2.9\times 10^{-4}$ \\ 
$|c_{d_1}^{YZ}|$ & $7.2\times 10^{-4}$ & $6.4\times 10^{-4}$ & $2.9\times 10^{-4}$ \\ 
$|c_{d_1}^{XX}-c_{d_1}^{YY}|$ & $1.3\times 10^{-3}$ & $1.2\times 10^{-3}$ & $5.5\times 10^{-4}$ \\ \hline
$|d_{u_1}^{XY}|$ & $8.3\times 10^{-3}$ & $7.5\times 10^{-3}$ & $3.4\times 10^{-3}$ \\ 
$|d_{u_1}^{XZ}|$ & $2.4\times 10^{-2}$ & $2.1\times 10^{-2}$ & $9.5\times 10^{-3}$ \\ 
$|d_{u_1}^{YZ}|$ & $2.4\times 10^{-2}$ & $2.1\times 10^{-2}$ & $9.5\times 10^{-3}$ \\ 
$|d_{u_1}^{XX}-d_{u_1}^{YY}|$ & $4.4\times 10^{-2}$ & $4.0\times 10^{-2}$ & $1.8\times 10^{-2}$ \\ \hline\hline
$|c_{u_2}^{XY}|$ & $1.5\times 10^{-3}$ & $1.4\times 10^{-3}$ & $6.2\times 10^{-4}$ \\ 
$|c_{u_2}^{XZ}|$ & $4.3\times 10^{-3}$ & $3.8\times 10^{-3}$ & $1.7\times 10^{-3}$ \\ 
$|c_{u_2}^{YZ}|$ & $4.3\times 10^{-3}$ & $3.8\times 10^{-3}$ & $1.7\times 10^{-3}$ \\ 
$|c_{u_2}^{XX}-c_{u_2}^{YY}|$ & $8.0\times 10^{-3}$ & $7.2\times 10^{-3}$ & $3.3\times 10^{-3}$ \\ \hline
$|c_{d_2}^{XY}|$ & $2.4\times 10^{-3}$ & $2.2\times 10^{-3}$ & $9.8\times 10^{-4}$ \\ 
$|c_{d_2}^{XZ}|$ & $6.8\times 10^{-3}$ & $5.9\times 10^{-3}$ & $2.7\times 10^{-3}$ \\ 
$|c_{d_2}^{YZ}|$ & $6.8\times 10^{-3}$ & $6.0\times 10^{-3}$ & $2.7\times 10^{-3}$ \\ 
$|c_{d_2}^{XX}-c_{d_2}^{YY}|$ & $1.3\times 10^{-2}$ & $1.1\times 10^{-2}$ & $5.2\times 10^{-3}$ \\ \hline
$|d_{u_2}^{XY}|$ & $2.9\times 10^{-1}$ & $2.7\times 10^{-1}$ & $1.2\times 10^{-1}$ \\ 
$|d_{u_2}^{XZ}|$ & $8.4\times 10^{-1}$ & $7.3\times 10^{-1}$ & $3.4\times 10^{-1}$ \\ 
$|d_{u_2}^{YZ}|$ & $8.4\times 10^{-1}$ & $7.4\times 10^{-1}$ & $3.4\times 10^{-1}$ \\ 
$|d_{u_2}^{XX}-d_{u_2}^{YY}|$ & $-$ & $-$ & $6.4\times 10^{-1}$ \\ \hline\hline
\end{tabular}
\caption{Expected best constraints of the time-dependent coefficients in the ($c_{u_i}$, $c_{d_i}$, $d_{u_i}$) basis. See the caption in table~\ref{tab:Q171} for further details. 
\label{tab:Q172}}
\end{table}

\begin{table}[t]
\centering
\begin{tabular}{|c||c|c|c||c|c|}\hline
\multirow{2}{*}{coefficient} & 
\multicolumn{3}{c||}{$
[\frac{d\sigma}{dQ}]_{Q=m_Z}
$} & \multicolumn{2}{c|}{$
[\frac{d\sigma}{dQ}]_{Q=m_Z} 
/
[\frac{d\sigma}{dQ}]_{Q=17.5 \; {\rm GeV}}
$}  \\ \cline{2-6}
 & nothing & $\delta_{\rm lumi}$ & $\delta_{\rm lumi}$, $\delta_{\rm sel}$ & $\delta_{\rm th}$, $\delta_{\rm lumi}$ & $\delta_{\rm th}$,  $\delta_{\rm lumi}$, $\delta_{\rm sel}$ \\ \hline
$|c_{U_1}^{XY}|$ & $2.4\times 10^{-3}$ & $6.8\times 10^{-4}$ & $3.3\times 10^{-4}$ & $4.3\times 10^{-5}$ & $2.0\times 10^{-5}$ \\ 
$|c_{U_1}^{XZ}|$ & $2.9\times 10^{-2}$ & $8.1\times 10^{-3}$ & $3.9\times 10^{-3}$ & $1.2\times 10^{-4}$ & $5.5\times 10^{-5}$ \\ 
$|c_{U_1}^{YZ}|$ & $2.8\times 10^{-2}$ & $8.1\times 10^{-3}$ & $3.9\times 10^{-3}$ & $1.2\times 10^{-4}$ & $5.4\times 10^{-5}$ \\ 
$|c_{U_1}^{XX}-c_{U_1}^{YY}|$ & $4.2\times 10^{-2}$ & $1.2\times 10^{-2}$ & $5.7\times 10^{-3}$ & $2.3\times 10^{-4}$ & $1.0\times 10^{-4}$ \\ \hline
$|c_{D_1}^{XY}|$ & $2.4\times 10^{-2}$ & $6.7\times 10^{-3}$ & $3.3\times 10^{-3}$ & $4.2\times 10^{-4}$ & $1.9\times 10^{-4}$ \\ 
$|c_{D_1}^{XZ}|$ & $2.9\times 10^{-1}$ & $8.0\times 10^{-2}$ & $3.9\times 10^{-2}$ & $1.2\times 10^{-3}$ & $5.4\times 10^{-4}$ \\ 
$|c_{D_1}^{YZ}|$ & $2.8\times 10^{-1}$ & $8.0\times 10^{-2}$ & $3.8\times 10^{-2}$ & $1.2\times 10^{-3}$ & $5.3\times 10^{-4}$ \\ 
$|c_{D_1}^{XX}-c_{D_1}^{YY}|$ & $4.2\times 10^{-1}$ & $1.2\times 10^{-1}$ & $5.7\times 10^{-2}$ & $2.2\times 10^{-3}$ & $1.0\times 10^{-3}$ \\ \hline
$|c_{Q_1}^{XY}|$ & $3.0\times 10^{-4}$ & $8.4\times 10^{-5}$ & $4.1\times 10^{-5}$ & $4.2\times 10^{-5}$ & $1.9\times 10^{-5}$ \\ 
$|c_{Q_1}^{XZ}|$ & $3.6\times 10^{-3}$ & $1.0\times 10^{-3}$ & $4.9\times 10^{-4}$ & $1.2\times 10^{-4}$ & $5.4\times 10^{-5}$ \\ 
$|c_{Q_1}^{YZ}|$ & $3.5\times 10^{-3}$ & $1.0\times 10^{-3}$ & $4.8\times 10^{-4}$ & $1.2\times 10^{-4}$ & $5.2\times 10^{-5}$ \\ 
$|c_{Q_1}^{XX}-c_{Q_1}^{YY}|$ & $5.2\times 10^{-3}$ & $1.5\times 10^{-3}$ & $7.1\times 10^{-4}$ & $2.2\times 10^{-4}$ & $1.0\times 10^{-4}$ \\ \hline\hline
$|c_{U_2}^{XY}|$ & $1.1\times 10^{-1}$ & $3.1\times 10^{-2}$ & $1.5\times 10^{-2}$ & $2.5\times 10^{-3}$ & $1.2\times 10^{-3}$ \\ 
$|c_{U_2}^{XZ}|$ & $-$ & $3.6\times 10^{-1}$ & $1.8\times 10^{-1}$ & $7.0\times 10^{-3}$ & $3.2\times 10^{-3}$ \\ 
$|c_{U_2}^{YZ}|$ & $-$ & $3.6\times 10^{-1}$ & $1.7\times 10^{-1}$ & $6.9\times 10^{-3}$ & $3.1\times 10^{-3}$ \\ 
$|c_{U_2}^{XX}-c_{U_2}^{YY}|$ & $-$ & $5.3\times 10^{-1}$ & $2.6\times 10^{-1}$ & $1.3\times 10^{-2}$ & $6.1\times 10^{-3}$ \\ \hline
$|c_{D_2}^{XY}|$ & $2.1\times 10^{-1}$ & $5.9\times 10^{-2}$ & $2.9\times 10^{-2}$ & $3.9\times 10^{-3}$ & $1.8\times 10^{-3}$ \\ 
$|c_{D_2}^{XZ}|$ & $-$ & $7.0\times 10^{-1}$ & $3.4\times 10^{-1}$ & $1.1\times 10^{-2}$ & $5.1\times 10^{-3}$ \\ 
$|c_{D_2}^{YZ}|$ & $-$ & $7.0\times 10^{-1}$ & $3.4\times 10^{-1}$ & $1.1\times 10^{-2}$ & $4.9\times 10^{-3}$ \\ 
$|c_{D_2}^{XX}-c_{D_2}^{YY}|$ & $-$ & $-$ & $5.0\times 10^{-1}$ & $2.1\times 10^{-2}$ & $9.7\times 10^{-3}$ \\ \hline
$|c_{Q_2}^{XY}|$ & $5.3\times 10^{-3}$ & $1.5\times 10^{-3}$ & $7.2\times 10^{-4}$ & $1.8\times 10^{-3}$ & $8.3\times 10^{-4}$ \\ 
$|c_{Q_2}^{XZ}|$ & $6.3\times 10^{-2}$ & $1.8\times 10^{-2}$ & $8.6\times 10^{-3}$ & $5.1\times 10^{-3}$ & $2.3\times 10^{-3}$ \\ 
$|c_{Q_2}^{YZ}|$ & $6.2\times 10^{-2}$ & $1.8\times 10^{-2}$ & $8.4\times 10^{-3}$ & $5.0\times 10^{-3}$ & $2.3\times 10^{-3}$ \\ 
$|c_{Q_2}^{XX}-c_{Q_2}^{YY}|$ & $9.2\times 10^{-2}$ & $2.6\times 10^{-2}$ & $1.3\times 10^{-2}$ & $9.6\times 10^{-3}$ & $4.4\times 10^{-3}$ \\ \hline\hline
\end{tabular}
\caption{Expected best constraints on the time-dependent coefficients in the ($c_{U_i}$, $c_{D_i}$, $c_{Q_i}$)  basis for $Q=m_Z$ and $Q=17.5\; {\rm GeV}$. See the caption in table~\ref{tab:Q171} for further details.
\label{tab:Zpole1}}
\end{table}

\begin{table}[t]
\centering
\begin{tabular}{|c||c|c|c||c|c|}\hline
\multirow{2}{*}{coefficient} & 
\multicolumn{3}{c||}{$
[\frac{d\sigma}{dQ}]_{Q=m_Z}
$} & \multicolumn{2}{c|}{$
[\frac{d\sigma}{dQ}]_{Q=m_Z} 
/
[\frac{d\sigma}{dQ}]_{Q=17.5 \; {\rm GeV}}
$}  \\ \cline{2-6}
 & nothing & $\delta_{\rm lumi}$ & $\delta_{\rm lumi}$, $\delta_{\rm sel}$ & $\delta_{\rm th}$, $\delta_{\rm lumi}$ & $\delta_{\rm th}$, $\delta_{\rm lumi}$, $\delta_{\rm sel}$ \\ \hline
$|c_{u_1}^{XY}|$ & $2.7\times 10^{-4}$ & $7.5\times 10^{-5}$ & $3.6\times 10^{-5}$ & $2.2\times 10^{-5}$ & $1.0\times 10^{-5}$ \\ 
$|c_{u_1}^{XZ}|$ & $3.2\times 10^{-3}$ & $9.0\times 10^{-4}$ & $4.3\times 10^{-4}$ & $6.1\times 10^{-5}$ & $2.8\times 10^{-5}$ \\ 
$|c_{u_1}^{YZ}|$ & $3.2\times 10^{-3}$ & $9.0\times 10^{-4}$ & $4.3\times 10^{-4}$ & $6.2\times 10^{-5}$ & $2.8\times 10^{-5}$ \\ 
$|c_{u_1}^{XX}-c_{u_1}^{YY}|$ & $4.6\times 10^{-3}$ & $1.3\times 10^{-3}$ & $6.3\times 10^{-4}$ & $1.2\times 10^{-4}$ & $5.4\times 10^{-5}$ \\ \hline
$|c_{d_1}^{XY}|$ & $1.2\times 10^{-2}$ & $3.3\times 10^{-3}$ & $1.6\times 10^{-3}$ & $2.1\times 10^{-4}$ & $9.7\times 10^{-5}$ \\ 
$|c_{d_1}^{XZ}|$ & $1.4\times 10^{-1}$ & $4.0\times 10^{-2}$ & $1.9\times 10^{-2}$ & $5.8\times 10^{-4}$ & $2.7\times 10^{-4}$ \\ 
$|c_{d_1}^{YZ}|$ & $1.4\times 10^{-1}$ & $4.0\times 10^{-2}$ & $1.9\times 10^{-2}$ & $5.9\times 10^{-4}$ & $2.7\times 10^{-4}$ \\ 
$|c_{d_1}^{XX}-c_{d_1}^{YY}|$ & $2.0\times 10^{-1}$ & $5.8\times 10^{-2}$ & $2.8\times 10^{-2}$ & $1.1\times 10^{-3}$ & $5.1\times 10^{-4}$ \\ \hline
$|d_{u_1}^{XY}|$ & $3.4\times 10^{-4}$ & $9.6\times 10^{-5}$ & $4.6\times 10^{-5}$ & $5.8\times 10^{-4}$ & $2.7\times 10^{-4}$ \\ 
$|d_{u_1}^{XZ}|$ & $4.1\times 10^{-3}$ & $1.2\times 10^{-3}$ & $5.5\times 10^{-4}$ & $1.6\times 10^{-3}$ & $7.4\times 10^{-4}$ \\ 
$|d_{u_1}^{YZ}|$ & $4.1\times 10^{-3}$ & $1.2\times 10^{-3}$ & $5.5\times 10^{-4}$ & $1.6\times 10^{-3}$ & $7.3\times 10^{-4}$ \\ 
$|d_{u_1}^{XX}-d_{u_1}^{YY}|$ & $5.9\times 10^{-3}$ & $1.7\times 10^{-3}$ & $8.0\times 10^{-4}$ & $3.1\times 10^{-3}$ & $1.4\times 10^{-3}$ \\ \hline\hline
$|c_{u_2}^{XY}|$ & $5.1\times 10^{-3}$ & $1.4\times 10^{-3}$ & $6.9\times 10^{-4}$ & $1.4\times 10^{-3}$ & $6.6\times 10^{-4}$ \\ 
$|c_{u_2}^{XZ}|$ & $6.1\times 10^{-2}$ & $1.7\times 10^{-2}$ & $8.3\times 10^{-3}$ & $3.9\times 10^{-3}$ & $1.8\times 10^{-3}$ \\ 
$|c_{u_2}^{YZ}|$ & $6.1\times 10^{-2}$ & $1.7\times 10^{-2}$ & $8.2\times 10^{-3}$ & $4.0\times 10^{-3}$ & $1.8\times 10^{-3}$ \\ 
$|c_{u_2}^{XX}-c_{u_2}^{YY}|$ & $8.8\times 10^{-2}$ & $2.5\times 10^{-2}$ & $1.2\times 10^{-2}$ & $7.6\times 10^{-3}$ & $3.5\times 10^{-3}$ \\ \hline
$|c_{d_2}^{XY}|$ & $1.0\times 10^{-1}$ & $2.9\times 10^{-2}$ & $1.4\times 10^{-2}$ & $2.0\times 10^{-3}$ & $9.1\times 10^{-4}$ \\ 
$|c_{d_2}^{XZ}|$ & $-$ & $3.5\times 10^{-1}$ & $1.7\times 10^{-1}$ & $5.5\times 10^{-3}$ & $2.5\times 10^{-3}$ \\ 
$|c_{d_2}^{YZ}|$ & $-$ & $3.5\times 10^{-1}$ & $1.7\times 10^{-1}$ & $5.5\times 10^{-3}$ & $2.5\times 10^{-3}$ \\ 
$|c_{d_2}^{XX}-c_{d_2}^{YY}|$ & $-$ & $5.1\times 10^{-1}$ & $2.4\times 10^{-1}$ & $1.1\times 10^{-2}$ & $4.8\times 10^{-3}$ \\ \hline
$|d_{u_2}^{XY}|$ & $5.6\times 10^{-3}$ & $1.6\times 10^{-3}$ & $7.6\times 10^{-4}$ & $1.0\times 10^{-2}$ & $4.6\times 10^{-3}$ \\ 
$|d_{u_2}^{XZ}|$ & $6.7\times 10^{-2}$ & $1.9\times 10^{-2}$ & $9.1\times 10^{-3}$ & $2.8\times 10^{-2}$ & $1.3\times 10^{-2}$ \\ 
$|d_{u_2}^{YZ}|$ & $6.7\times 10^{-2}$ & $1.9\times 10^{-2}$ & $9.1\times 10^{-3}$ & $2.8\times 10^{-2}$ & $1.3\times 10^{-2}$ \\ 
$|d_{u_2}^{XX}-d_{u_2}^{YY}|$ & $9.7\times 10^{-2}$ & $2.8\times 10^{-2}$ & $1.3\times 10^{-2}$ & $5.3\times 10^{-2}$ & $2.4\times 10^{-2}$ \\ \hline\hline
\end{tabular}
\caption{Expected best constraints of the time-dependent coefficients in the ($c_{u_i}$, $c_{d_i}$, $d_{u_i}$) basis for $Q=m_Z$ and $Q=17.5\; {\rm GeV}$. See the captions in table~\ref{tab:Q171} and table~\ref{tab:Zpole1} for further details. 
\label{tab:Zpole2}}
\end{table}

The remaining coefficient combinations $XZ, YZ, XY, XX-YY$ in the Sun-centered frame produce sidereal oscillations as a function of $T_\oplus$. These effects average to zero using data taken over long periods of time, which is typically the case for collider cross-section measurements. Thus, without event timestamps, existing SM measurements cannot be used to constrain time-dependent effects---instead, simulated or expected constraints can be placed. The procedure adopted to calculate the expected bounds has been detailed in several previous analyses~\cite{Kostelecky:2016pyx, Lunghi:2018uwj, Kostelecky:2019fse}. The idea is to simulate the outcome of binning the data in sidereal time and to perform measurements in each time bin. For those coefficients that generate dependence on sidereal time without affecting the time-averaged cross section, one can show that sources of systematic experimental uncertainties 100\% correlated across time bins do not impact the extracted bounds. 

The dominant source of systematic uncertainties in the measurements at low $Q^2$ presented in ref.~\cite{Sirunyan:2018owv} originate from theory ($\delta_{\text{th}})$, luminosity ($\delta_{\text{lumi}})$, and lepton selection efficiency ($\delta_{\text{sel}})$. While theory errors are 100\% correlated across sidereal bins, the latter two are more complex. In fact, the stochastic components of uncertainties associated with luminosity and selection efficiency are actually uncorrelated in time bins. We present the three limiting cases in which the luminosity and selection efficiency are considered either uncorrelated or 100\% correlated. In all cases, the theoretical uncertainty is taken to be fully correlated.
The measurement of the $Z$-pole DY cross section presented in ref.~\cite{Sirunyan:2019bzr} has subdominant statistical uncertainties and among the systematic errors most likely to be partially correlated in sidereal time we focus on the luminosity and selection efficiency. Both uncertainties have a stochastic component that is difficult to estimate without a detailed experimental analysis. We present three scenarios in which we assume that no errors (nothing), luminosity, and both luminosity and selection efficiency are correlated between sidereal bins. 

Finally, we present bounds that can be extracted from an analysis of the cross sections at $Q = m_Z$ and $Q = 17.5\; {\rm GeV}$. This ratio has the considerable advantage that luminosity uncertainties of numerator and denominator exactly cancel because they are 100\% correlated. This is true to a certain extent for systematic errors due to selection efficiency. Accordingly, we consider the two cases in which either $(\delta_{\rm th},\delta_{\rm lumi})$ or $(\delta_{\rm th},\delta_{\rm lumi},\delta_{\rm sel})$ are correlated. While the first scenario is certainly correct, we have higher confidence in the bin-to-bin correlation of selection efficiency for the ratio rather than for the individual cross sections. In conclusion, we expect that the bounds corresponding to the second scenario are on more solid ground than the most aggressive bounds we obtain from the cross sections at $Q=m_Z$ and $Q=17.5\; {\rm GeV}$. 

The expected upper bounds that we obtain from the $Q = 17.5\; {\rm GeV}$ measurement are presented in tables~\ref{tab:Q171} and \ref{tab:Q172} for the two bases, respectively. As expected, from table~\ref{tab:Q172} it is observed that the bounds on $d_{u_1}^{JK}$ are more than an order of magnitude worse than on $c_{u_1}^{JK}$. This also underscores how the bounds in table~\ref{tab:Q171} are highly correlated.
The corresponding results for the $Z$-pole measurement and the ratio of cross sections are presented in tables~\ref{tab:Zpole1} and \ref{tab:Zpole2}. We observe that considering $Z$-pole measurements alone, the expected bounds on the $d$ coefficients are actually stronger than those for the $c$ coefficients. Interestingly, this is not the case when considering the ratio of cross sections. The reason is that the $d_{u_i}$ dependence of the cross sections at $Q=m_Z$ and $Q = 17.5\; {\rm GeV}$ cancels in the ratio.

\section{Conclusions and future outlook}
In this work, we have extended the framework developed in \cite{Kostelecky:2019fse} to describe spin-dependent effects coupled through the production of $Z$ bosons. Real and estimated bounds on CPT-even and dimensionless quark-sector coefficients for Lorentz violation for the first two generations of quarks are placed using data taken at the LHC. Bounds generically range from the $10^{-1}$-$10^{-5}$ level, with the most sensitivity for the light quarks given their greater PDFs magnitudes at low $x$. These results highlight encouraging prospects for expanding the space of constraints on numerous quark-sector Lorentz-violating effects. Future studies incorporating mass effects from heavy quarks, charged-current processes, polarized processes, and effects from radiative corrections and nonabelian gauge fields, will significantly extend the reach for detecting potential violations of CPT and Lorentz invariance in existing and future collider experiments.
\section{Acknowledgments}
We thank V. Alan Kosteleck\'y for useful discussions. This work was supported in part
by the U.S.\ Department of Energy under grants No. {DE}-SC0010120,
No. DE-AC05-06OR23177, No. DE-FG02-87ER4036, and by the Indiana Space Grant Consortium.

\bibliographystyle{JHEP} 
\bibliography{draft}

\end{document}